# Elemental and molecular abundances in comet 67P/Churyumov-Gerasimenko

Martin Rubin[1,*], Kathrin Altwegg[1], Hans Balsiger[1], Jean-Jacques Berthelier[2], Michael R. Combi[3], Johan De Keyser[4], Maria Drozdovskaya[5], Björn Fiethe[6], Stephen A. Fuselier[7,8], Sébastien Gasc[1], Tamas I. Gombosi[3], Nora Hänni[1], Kenneth C. Hansen[3], Urs Mall[9], Henri Rème[10], Isaac R. H. G. Schroeder[1], Markus Schuhmann[1], Thierry Sémon[1], Jack H. Waite[7], Susanne F. Wampfler[5], and Peter Wurz[1,5]

[1]Physikalisches Institut, University of Bern, Sidlerstrasse 5, CH-3012 Bern, Switzerland.
[2]Laboratoire Atmosphères, Milieux, Observations Spatiales, Institut Pierre Simon Laplace, CNRS, Université Pierre et Marie Curie, 4 Avenue de Neptune, 94100, Saint-Maur, France.
[3]Department of Climate and Space Sciences and Engineering, University of Michigan, 2455 Hayward, Ann Arbor, MI 48109, USA.
[4]Koninklijk Belgisch Instituut voor Ruimte-Aeronomie - Institut Royal Belge d'Aéronomie Spatiale, Ringlaan 3, B-1180 Brussels, Belgium.
[5]Center for Space and Habitability, University of Bern, Gesellschaftsstrasse 6, CH-3012 Bern, Switzerland.
[6]Institute of Computer and Network Engineering (IDA), TU Braunschweig, Hans-Sommer-Straße 66, D-38106 Braunschweig, Germany.
[7]Space Science Directorate, Southwest Research Institute, 6220 Culebra Rd., San Antonio, TX 78228, USA.
[8]University of Texas at San Antonio, San Antonio, TX USA
[9]Max-Planck-Institut für Sonnensystemforschung, Justus-von-Liebig-Weg 3, 37077 Göttingen, Germany.
[10]Institut de Recherche en Astrophysique et Planétologie, CNRS, Université Paul Sabatier, Observatoire Midi-Pyrénées, 9 Avenue du Colonel Roche, 31028 Toulouse Cedex 4, France.

## Abstract

Comets are considered to be some of the most pristine and unprocessed solar system objects accessible to in-situ exploration. Investigating their molecular and elemental composition takes us on a journey back to the early period of our solar system and possibly even further. In this work, we deduce the bulk abundances of the major volatile species in comet 67P/Churyumov-Gerasimenko, the target of the European Space Agency's Rosetta mission. The basis are measurements obtained with the ROSINA instrument suite on board the Rosetta orbiter during a suitable period of high outgassing near perihelion. The results are combined with both gas and dust composition measurements published in the literature. This provides an integrated inventory of the major elements present in the nucleus of 67P/Churyumov-Gerasimenko. Similar to comet 1P/Halley, which was visited by ESA's Giotto spacecraft in 1986, comet 67P/Churyumov-Gerasimenko also shows near-solar abundances of oxygen and carbon, whereas hydrogen and nitrogen are depleted compared to solar. Still, the degree of devolatilization is lower than that of inner solar system objects, including meteorites and the Earth. This supports the idea that comets are among the most pristine objects in our solar system.

* E-mail: martin.rubin@space.unibe.ch

## Keywords

comets: general
comets: individual: 67P/Churyumov-Gerasimenko

# 1. Introduction

Comets were formed by the accretion of condensed matter originating from various nucleosynthetic processes (Geiss 1987). As such, they contain sizeable amounts of volatile species in the form of ices next to lesser volatile organics all the way to refractory material. The origin of the material is still debated. The relative abundances of volatile materials, in particular, indicate an interstellar medium (ISM) origin (Bockelée-Morvan *et al.* 2000), where chemistry in the icy mantles of grains played an important role (Herbst & van Dishoeck 2009). Cold temperatures in the ISM also affect isotopic ratios, in particular the deuteration of molecules (Furuya *et al.* 2016), and hence several isotopic ratios in the volatile species also point to an origin in the ISM (Bockelée-Morvan *et al.* 2015).

On the other hand, a large number of high temperature condensates that were processed close to the Sun or by shocks in the disk have been identified in the samples returned by NASA's Stardust mission to comet Wild 2 (McKeegan *et al.* 2006, Brownlee *et al.* 2006). Therefore, looking at comets as a whole, multiple processes must have been at work in forming the material they are made from and distributing it in the solar system. Molecular and elemental abundances thus contain crucial clues to the origin of the cometary material and the subsequent processing during or even before the formation of the solar system.

Addressing these questions was one of the main goals of the European Space Agency's (ESA) Rosetta mission to the Jupiter Family Comet (JFC) 67P/Churyumov-Gerasimenko (Schwehm & Schulz 1999, Glassmeier *et al.* 2007). Rosetta followed the comet for more than two years starting in August 2014 to the end of September 2016. The spacecraft carried a suite of instruments dedicated to a close investigation, including the lander Philae which was successfully deployed on the surface of the nucleus in November 2014 (Bibring *et al.* 2015).

Compositional measurements of comet 67P/Churyumov-Gerasimenko (hereafter 67P/C-G) were carried out with several instruments. Volatile species were measured by Alice (Feldman *et al.* 2015), VIRTIS (Visible and Thermal Infrared Thermal Imaging Spectrometer; Bockelée-Morvan *et al.* (2016)), ROSINA (Rosetta Orbiter Spectrometer for Ion and Neutral Analysis; Le Roy et al. (2015)), and OSIRIS (Optical, Spectroscopic, and Infrared Remote Imaging System; Bodewits *et al.* (2016)) on board the Rosetta spacecraft and by Ptolemy (Wright *et al.* 2015) and COSAC (Cometary Sampling and Composition; Goesmann *et al.* (2015)) on board the lander Philae.

ROSINA was the mass spectrometer suite on board Rosetta dedicated to the in-situ measurement of the volatile material emanating from the comet's nucleus (Balsiger *et al.* 2007). ROSINA consisted of the Double Focusing Mass Spectrometer (DFMS), the Reflectron-type Time-Of-Flight mass spectrometer (RTOF), and the Comet Pressure Sensor (COPS). The two mass spectrometers complemented each other with high time-resolution, mass-resolution, sensitivity, and dynamic range.

ROSINA observations revealed a plethora of volatiles released from the nucleus already at a heliocentric distance beyond 3 au (Le Roy et al. 2015). The most abundant volatiles, $H_2O$, $CO_2$, and CO, were shown to exhibit a pronounced heterogeneous distribution in the coma (Hässig *et al.* 2015, Hoang *et al.* 2017). Highly volatile $O_2$, found in similar quantity as $CO_2$ and CO, showed a good correlation to $H_2O$ (Bieler et al. 2015, Gasc *et al.* 2017a). Numerical models of the most abundant volatiles of the comet were then used to quantify the activity and its distribution on the surface (Hansen *et al.* 2016, Fougere *et al.* 2016b, Läuter *et al.* 2019, Marschall *et al.* 2017, Zakharov *et al.* 2018). In addition, species with (much) lower abundances, including a host of S-bearing molecules (Calmonte et al. 2016), hydrocarbons and oxygenated compounds (Schuhmann et al. 2019, Schuhmann et al. in preparation), noble gases (Balsiger et al. 2015, Marty *et al.* 2017, Rubin et al. 2018), hydrogen halides (Dhooghe et al. 2017), and $N_2$ (Rubin et al. 2015a) were detected during various phases of the mission.

The abundances in the refractory elements in the dust of 67P/C-G were further measured by the Comet Secondary Ion Mass Spectrometer (COSIMA; Kissel *et al.* (2007)), and presented by Bardyn et al. (2017). Grains collected on COSIMA's exposed target were bombarded with an 8 keV indium ion beam and the ions sputtered off the dust were detected by a micro-channel plate (MCP) detector after mass-separation in a time-of-flight section. Given the collector temperature of around 283 K, COSIMA was most sensitive to the refractory material and the less volatile organics (Fray et al. 2016) that remained on the collectors for up to several days or even weeks without evaporating.

In this paper, we report on the deduced bulk abundances of the volatile species in comet 67P/C-G. For this purpose, we used ROSINA data to derive relative abundances of $CO_2$, CO, $O_2$, $NH_3$, HCN, HCNO, $CH_3NO$, $CH_3CN$, $HC_3N$ and their isomers with respect to $H_2O$ for a suitable period near perihelion. These results were then combined with relative abundances derived from ROSINA data in the same fashion from published literature. The following section presents the two ROSINA instruments used in this work, DFMS and COPS, and the analytical technique used to derive relative abundances from the measurements. The next section discusses the measurement period selected for deducing the bulk abundances of the volatile material in the comet's nucleus. Afterwards we present a collection of our results together with previously published bulk abundances derived from ROSINA observations. The elemental abundances in 67P/C-G's ices follow from these data, and when combined with COSIMA results, the elemental ratios in 67P/C-G's nucleus as well. The last section discusses our findings and compares them with other objects in our solar system.

## 2. Instrument description and data reduction

DFMS was a high-resolution, high-sensitivity mass spectrometer built in double-focusing configuration (Mattauch & Herzog 1934). Neutral gas particles from the comet entered DFMS' ion source, where they were ionized and partially fragmented by 45 eV electrons emitted by a hot filament. After ionization, the ions were accelerated and deflected by 90° in the electric field of the curved electrostatic analyser. After another deflection by 60° in a permanent magnet, they passed the zoom optics, where the ion beam was widened before it impacted the MCP detector. The resulting mass resolution was m/Δm = 3000 for a mass/charge ratio of 28 u/e at 1% of the peak height (Balsiger et al. 2007). For mass/charge ratios of m/Q ≥ 70 u/e, an additional post-acceleration potential was applied in front of the detector to increase the detection efficiency for the heavier species. The avalanche of electrons produced by each impact on the MCP was collected on the position-sensitive Linear Electron Detector Array's (LEDA) two parallel rows of 512 pixels each, which were operated in analogue mode (Nevejans *et al.* 2002).

Each mass-to-charge ratio was measured separately in a sequential order. A suitable set of potentials would first be applied for a given mass-to-charge ratio and the voltage across the MCP adjusted to guarantee suitable amplification while avoiding saturation. This allowed variation of the gain between each measurement, resulting in a high dynamic range on the order of $10^{10}$. Setting the voltages took roughly 10 s per spectrum. Each mass would then be accorded 20 s of integration time before moving on to the next mass. Hence, a full scan in the typical mass range from 13 u/e to 100 u/e lasted approximately 45 min. This includes the two additional 18 u/e measurements performed at the beginning and end of each measurement mode to monitor changes in the activity of the comet.

DFMS was used to derive relative abundances of the individual species with respect to water. To do so, linear time-interpolations between adjacent measurements of the same species were performed and related to water, which was used as reference. In the aforementioned measurement technique, ions impacted the MCP with different velocities and energies. All Level-3 data available through ESA's Planetary Science Archive (PSA) thus contains corrections for the MCP detector's yield (relative to neon) as derived during calibrations. With $a_X$, the number of counts for a given species X interpolated to a given time, $S_X$, its calibrated sensitivity, $Y_X$, its yield, $W_X = S_X Y_X$, its yield-corrected sensitivity, and $f_{X\to X^+}$, the fragmentation branching of neutral species $X$ to its parent ion $X^+$, the relative abundance of, e.g., carbon dioxide with respect to water can be computed with:

$$r_{CO_2} = \frac{n_{CO_2}}{n_{H_2O}} = \frac{a_{CO_2}\,\cos(\vartheta_{H_2O})\;W_{H_2O}\,f_{H_2O\to H_2O^+}}{a_{H_2O}\,\cos(\vartheta_{CO_2})\;W_{CO_2}\,f_{CO_2\to CO_2{}^+}}\quad , \tag{1}$$

where $\vartheta_X$ corresponds to the off-nadir pointing angle during the time of the measurement of species $X$. Thus, $\cos(\vartheta_X)$ represents the geometrical cross-section of DFMS' ion source that is exposed to the comet. Data obtained at times when the spacecraft was pointing more than 30° off from the nadir direction have been excluded from the dataset. Additionally, considering that all species were measured within ~20 min of a water measurement and slew rates were slow, the $\cos(\vartheta_{H_2O})\,/\cos(\vartheta_X)$-correction tends to be small.

The case of, e.g., carbon monoxide is more complicated compared to the example above as the fragmentation of $CO_2$ and $H_2CO$ to $CO^+$ inside the ion source has to be taken into account by subtracting the corresponding signal from the fragmentation $f_{X\to Y^+}$ of neutral species $X$ to its fragment ion $Y^+$ as follows:

$$r_{CO} = \frac{n_{CO}}{n_{H_2O}} = \frac{\left(\frac{a_{CO}}{Y_{CO}} - \frac{a_{CO_2}}{Y_{CO_2}} \frac{f_{CO_2 \to CO^+}}{f_{CO_2 \to CO_2{}^+}} - \frac{a_{H_2CO}}{Y_{H_2CO}} \frac{f_{H_2CO \to CO^+}}{f_{H_2CO \to H_2CO^+}}\right) W_{H_2O}\, Y_{CO}\, f_{H_2O \to H_2O^+}}{a_{H_2O}\, W_{CO}\, f_{CO \to CO^+}} \quad . \tag{2}$$

For the sake of readability, we have omitted the off-nadir correction from this equation. Additional contributions to the $CO^+$ signal from fragmentations of larger molecules were treated similarly. The same approach was taken with other species similarly affected by contributions from higher mass molecules, such as formaldehyde and methanol's contribution to it ($CH_3OH + e^- \rightarrow H_2CO^+ + 2H + 2e^-$). Gasc *et al.* (2017b) presented a simplified version of these equations for the analysis of RTOF data given that the different species are measured contemporaneously in a time-of-flight instrument.

The relative abundances $r_X$ of the volatile species $X = \{H_2O, CO, CO_2, O_2, H_2CO, \dots\}$ can now be derived with respect to water. Details on the ROSINA calibrated sensitivities $S_X$ and fragmentation patterns $f_{X \to Y^+}$ can be obtained from PSA, along with the Level-3 ROSINA datasets. For species that were not calibrated due to their toxicity or corrosiveness, we have applied the empirical formula for the sensitivity derived by Calmonte (2015) and included fragmentation patterns obtained from reference spectra published by NIST, the National Institute of Standards and Technology (Kim *et al.* 2005). Sensitivity errors for the individual species are on the order of 15 - 20%. An additional error of 20% is applicable for the uncertainty of the detector's gain calibration as well as aging processes caused by the accumulated lifetime charge, which differs from one pixel to the next (cf. Schroeder *et al.* (2018)). Furthermore, counting statistics and, in the case of overlapping peaks, a fitting error have been included (cf. Marty et al. (2017), Rubin et al. (2018)). The DFMS calibration factors used in this work have been summarized in Table 1.

The COPS sensor contained two gauges, the Nude Gauge (NG) and the Ram Gauge. Here we used the NG, an open source where 150 eV electrons ionized the ambient neutral gas passing through the gauge. The ions formed inside the open volume were then accelerated toward a cathode and the ion current was measured after a suitable amplification. The ion current relative to the electron current is related to the density of the neutral gas inside the NG after application of the laboratory-derived calibration factors (Graf *et al.* 2004, Tzou 2017). The simplicity of the sensor makes it a reliable and stable monitor for the gas density of the comet in the vicinity of the spacecraft. No measurable change in calibration parameters has been observed for the laboratory twin of COPS throughout its operation time (Tzou 2017) despite heavy use.

COPS itself could not derive the composition of the gas. Therefore, COPS was calibrated for a 100% $N_2$ reference atmosphere (cf. Gasc et al. (2017b)). For other compositions, a set of correction factors has to be applied, as each species has its own ionization cross-section and hence a different sensitivity. The measured density is thus an $N_2$-equivalent and corresponds to the sum of the corrected partial densities

$$n_{COPS} = n_{N_2} = \sum_X \frac{n_X}{\beta_X} = n_{H_2O} \sum_X \frac{n_X}{n_{H_2O}\beta_X} \quad , \tag{3}$$

where $\beta_X$ is the correction factor for species $X$ with respect to $N_2$. This factor can be obtained from calibration or from the ratio of the total ionization cross-section of species $X$ with respect to $N_2$ for 150 eV electrons. The COPS correction factors used in this work can be found in Table 1. The density of water molecules can then be computed by using the relative abundance of species $X$ with respect to water, $r_X$, derived from DFMS

$$n_{\mathrm{H_2O}} = \frac{n_{\mathrm{COPS}}}{\sum_X \frac{n_X}{n_{\mathrm{H_2O}}\beta_X}} = \frac{n_{\mathrm{COPS}}}{\sum_X \frac{r_X}{\beta_X}} \quad . \tag{4}$$

Table 1: Calibration factors used for a subset of the volatile species detected at 67P/C-G, including the ones presented in this work.

| Species | Sensitivity [cm³] $W_X = S_X Y_X$ | Yield correction relative to neon $Y_X$ | COPS correction w.r.t. $N_2$ $\beta_X$ | Fragmentation (w.r.t. total ionization cross section) $f_{X \to Y^+}$ | |
|---|---|---|---|---|---|
| $H_2O$ | $2.04 \cdot 10^{-19}$ | 0.886 | 0.893 | $H_2O \to H_2O^+$ | 0.792 |
| $CO_2$ | $3.29 \cdot 10^{-19}$ | 2.141 | 0.704 | $CO_2 \to CO_2^+$<br>$CO_2 \to CO^+$ | 0.779<br>0.099 |
| $CO$ | $2.88 \cdot 10^{-19}$ | 1.420 | 0.952 | $CO \to CO^+$ | 0.964 |
| $O_2$ | $2.57 \cdot 10^{-19}$ | 1.623 | 0.990 | $O_2 \to O_2^+$ | 0.821 |
| $CH_4$ | $6.85 \cdot 10^{-19}$ | 0.790 | 0.654 | $CH_4 \to CH_4^+$ | 0.522 |
| $C_2H_6$ | $3.50 \cdot 10^{-19}$ | 1.526 | 0.380 | $C_2H_6 \to C_2H_6^+$ | 0.106 |
| $C_3H_8$ | $8.91 \cdot 10^{-19}$ | 2.144 | 0.262 | $C_3H_8 \to C_3H_8^+$ | 0.098 |
| $C_6H_6$ | $4.66 \cdot 10^{-18}$ | 2.624 | 0.172 | $C_6H_6 \to C_6H_6^+$ | 0.387 |
| $CH_3OH$ | $1.12 \cdot 10^{-18}$ | 1.624 | 0.541 | $CH_3OH \to H_2CO^+$<br>$CH_3OH \to CH_3OH^+$ | 0.030<br>0.130 |
| $C_2H_5OH$ | $6.99 \cdot 10^{-19}$ | 2.232 | 0.335 | $C_2H_5OH \to H_2CO^+$<br>$C_2H_5OH \to CH_3CHO^+$<br>$C_2H_5OH \to C_2H_5OH^+$ | 0.0083<br>0.0083<br>0.030 |
| $H_2CO$ | $4.86 \cdot 10^{-19}$ | 1.525 | 0.631 | $H_2CO \to H_2CO^+$ | 0.309 |
| $CH_3CHO$ | $6.90 \cdot 10^{-19}$ | 2.143 | 0.385 | $CH_3CHO \to H_2CO^+$<br>$CH_3CHO \to CH_3CHO^+$ | 0.0044<br>0.146 |
| $HCOOH$ | $4.72 \cdot 10^{-19}$ | 2.230 | 0.478 | $HCOOH \to H_2CO^+$<br>$HCOOH \to HCOOH^+$ | 0.0063<br>0.227 |
| $CH_3COOH$ | $6.16 \cdot 10^{-19}$ | 3.047 | 0.309 | $HCOOCH_3 \to H_2CO^+$<br>$HCOOCH_3 \to CH_3CHO^+$<br>$HCOOCH_3 \to HCOOCH_3^+$ | 0.0012<br>0.0074<br>0.222 |
| $NH_3$ | $3.83 \cdot 10^{-19}$ | 0.837 | 0.787 | $NH_3 \to NH_3^+$ | 0.475 |
| $HCN$ | $4.27 \cdot 10^{-19}$ | 1.367 | 0.645 | $HCN \to CN^+$<br>$HCN \to HCN^+$ | 0.133<br>0.792 |
| $HNCO$ | $3.51 \cdot 10^{-19}$ | 2.099 | 0.669 | $HNCO \to CN^+$<br>$HNCO \to HCN^+$<br>$HNCO \to CO^+$<br>$HNCO \to HNCO^+$ | 0.013<br>0.012<br>0.030<br>0.674 |
| $CH_3NO$ | $5.67 \cdot 10^{-19}$ | 2.186 | 0.402 | $CH_3NO \to CN^+$<br>$CH_3NO \to HCN^+$<br>$CH_3NO \to HNCO^+$<br>$CH_3NO \to NH_2CHO^+$ | 0.0053<br>0.031<br>0.045<br>0.411 |
| $CH_3CN$ | $6.27 \cdot 10^{-19}$ | 2.017 | 0.386 | $CH_3CN \to CH_3CN^+$ | 0.511 |
| $HC_3N$ | $6.02 \cdot 10^{-19}$ | 2.483 | 0.349 | $HC_3N \to HC_3N^+$ | 0.593 |
| $H_2S$ | $1.21 \cdot 10^{-19}$ | 1.717 | 0.455 | $H_2S \to H_2S^+$ | 0.574 |
| $OCS$ | $3.94 \cdot 10^{-19}$ | 3.044 | 0.532 | $OCS \to OCS^+$ | 0.528 |
| $SO_2$ | $3.68 \cdot 10^{-19}$ | 3.240 | 0.455 | $SO_2 \to SO_2^+$ | 0.575 |
| $CS_2$ | $2.68 \cdot 10^{-18}$ | 2.590 | 0.207 | $CS_2 \to CS_2^+$ | 0.664 |

The abundances of the other volatile species then follow directly from

$$n_X = r_X \, n_{\mathrm{H_2O}} \quad . \tag{5}$$

This technique provides a reliable way to derive accurate particle densities and relative abundances and was used throughout the ROSINA data analysis (cf. Gasc et al. (2017b)). It is independent of the aging processes of DFMS' MCP detector, provided that all species are, to the first order, affected in the same way by the decreasing MCP amplification over the course of the mission.

## 3. Measurements

ROSINA DFMS measured the relative abundances of many volatile species at the location of the Rosetta spacecraft. Besides water, the major volatile species were carbon dioxide ($CO_2$), carbon monoxide (CO), molecular oxygen ($O_2$), followed by methane ($CH_4$), ethane ($C_2H_6$), formaldehyde ($H_2CO$), and methanol ($CH_3OH$).

The ratios varied over the course of the mission and thus did not always reflect their abundance in the bulk ice of the nucleus. Figure 1 shows the ratios of HCN, HNCO, and $CH_3NO$ (including their isomers) with respect to $H_2O$ which are reported in this work. The measured ratios depend in a non-trivial way on the heliocentric distance, the orientation of 67P/C-G's spin axis, and Rosetta's location with respect to the nucleus (Gasc et al. 2017a, Hansen et al. 2016, Fougere *et al.* 2016a).

We thus selected the same measurement period from the inbound portion of the comet's orbit as Calmonte et al. (2016), namely May 2015, to derive the bulk abundances of $CO_2$, CO, $O_2$, $NH_3$, HCN, HCNO, $CH_3NO$, $CH_3CN$, $HC_3N$ and their isomers with respect to $H_2O$ in 67P/C-G. Calmonte et al. (2016) proposed this time period to be suitable to deduce bulk abundances of volatile species in the cometary ices. The reasons are: a) the high outgassing as the comet approaches perihelion and hence high erosion associated with providing fresh material from the comet's interior, b) little contamination from outbursts driven by highly volatile species that occurred close to perihelion in August 2015, and c) the favourable orbit for an in situ instrument such as ROSINA as Rosetta passed over the dayside of the nucleus and above the more active southern summer hemisphere. We hence regard the ratios measured in the coma during this period as representative of the bulk abundances in the ices of 67P/C-G. Figure 2 shows the details of Rosetta's location with respect to the nucleus. The grey bar at the bottom denotes the selected investigation period, when Rosetta was passing over the dayside (phase angles < 85°) of the active southern summer hemisphere (sub-spacecraft latitudes < 0°). The same period is also indicated in Figure 1 for comparison.

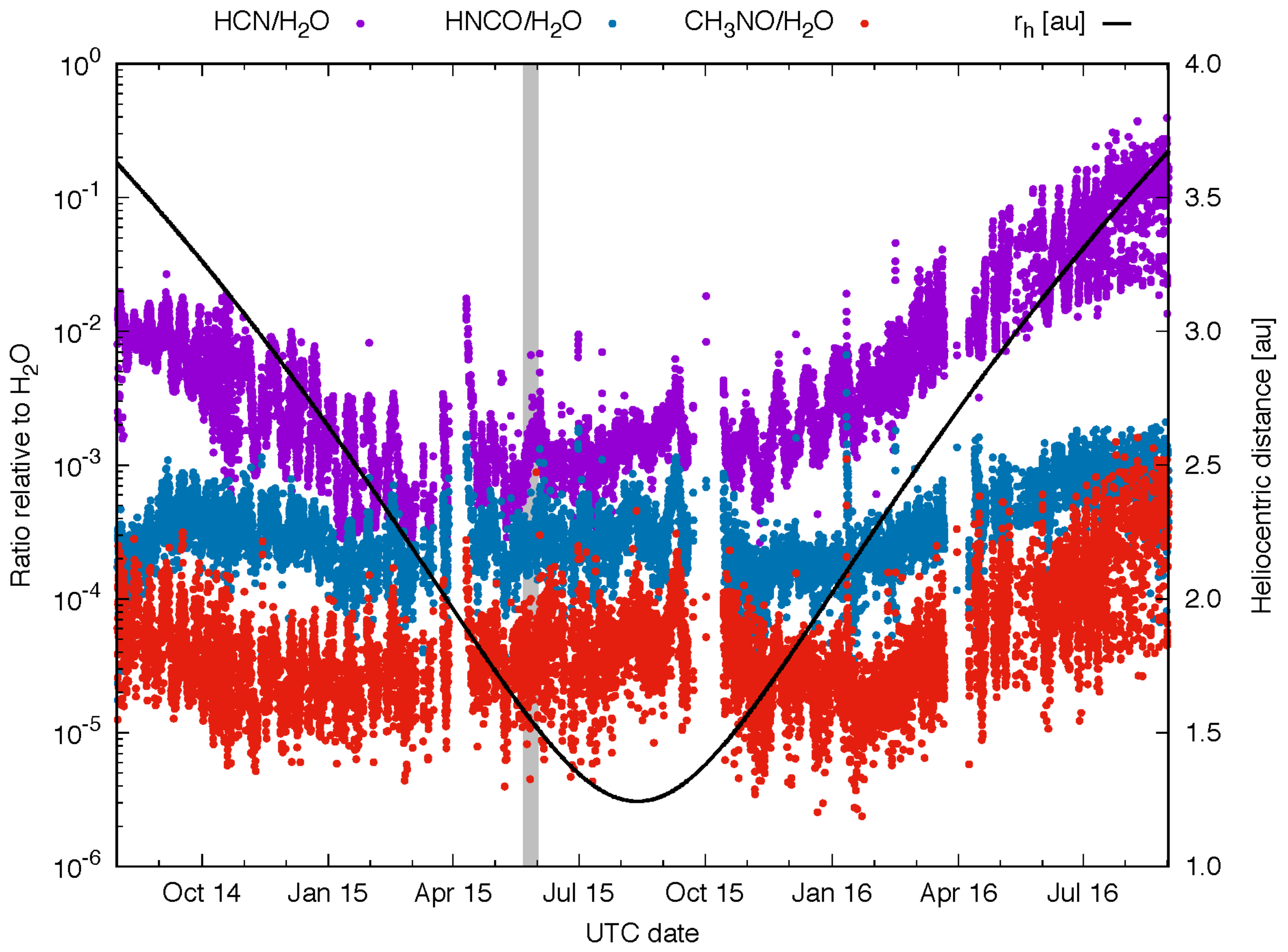

Figure 1: Relative abundances of HCN, HNCO, and $CH_3NO$ (and their isomers) with respect to $H_2O$ and the heliocentric distance (solid black line) throughout most of the mission, including the investigated period indicated in gray (cf. Figure 2).

The measured relative abundances with respect to water for the period investigated are shown in Figure 3. The horizontal lines represent the averaged ratios as indicated in the labels together with their corresponding 1-σ errors.

The same period in May 2015 has been analysed in previous works on the sulphur-bearing species (Calmonte et al. 2016), hydrocarbons and oxygenated compounds (Schuhmann et al. 2019, Schuhmann et al. in preparation), halogens (Dhooghe et al. 2017), noble gases, and $N_2$ (Rubin et al. 2018). We therefore refer to these works for the details and just include the corresponding relative bulk abundances in Table 2.

These results can be compared to measurements obtained earlier in the mission beyond 3.1 au. Le Roy et al. (2015) reported strong differences in the relative abundances with respect to water above the summer and winter hemispheres (reproduced in Table 2). The complex shape of the nucleus combined with the obliquity of the comet's rotation axis (Sierks *et al.* 2015) of 52° led to a pronounced seasonal outgassing pattern (Hässig et al. 2015). The deduced bulk abundances in Table 2 for the southern summer in May 2015 are within a factor of a few of those measured above the northern summer hemisphere before equinox (Le Roy et al. 2015). The northern hemisphere may either be enhanced in water or depleted in $CO_2$ (and other species of higher volatility). The May 2015 measurements, on the other hand, were obtained much closer to the Sun and hence at a higher relative water activity (cf. Gasc et al. (2017a)). Nevertheless, there are a few notable differences in both datasets, including the sulphur-bearing species SO, $SO_2$, and $S_2$, which may be partially associated with semi-volatile species in grains (Calmonte et al. 2016, Altwegg *et al.* 2016). Such grains are subject to higher sublimation rates due to increased grain temperatures at smaller heliocentric distances (Lien 1990), whereas at increased heliocentric distances their presence may remain hidden. This may also be the case for $CH_3NO$, $HC_3N$, and their corresponding isomers. The factor of 10 difference for $NH_3$, on the other hand, requires a different explanation and shall be addressed elsewhere.

DFMS measured density ratios in situ at the location of Rosetta; however, deriving nucleus outgassing flux ratios depends also on the velocities of the individual species. Analogous to Rubin et al. (2018), we have assumed collisional coupling of the gases in the near nucleus environment at small heliocentric distances, which in effect equilibrates the different outflow velocities. This is supported by numerical simulations (Tenishev *et al.* 2008). As a result, the measured density ratios at the spacecraft can be converted directly into the ratio of the outgassing fluxes from the cometary surface.

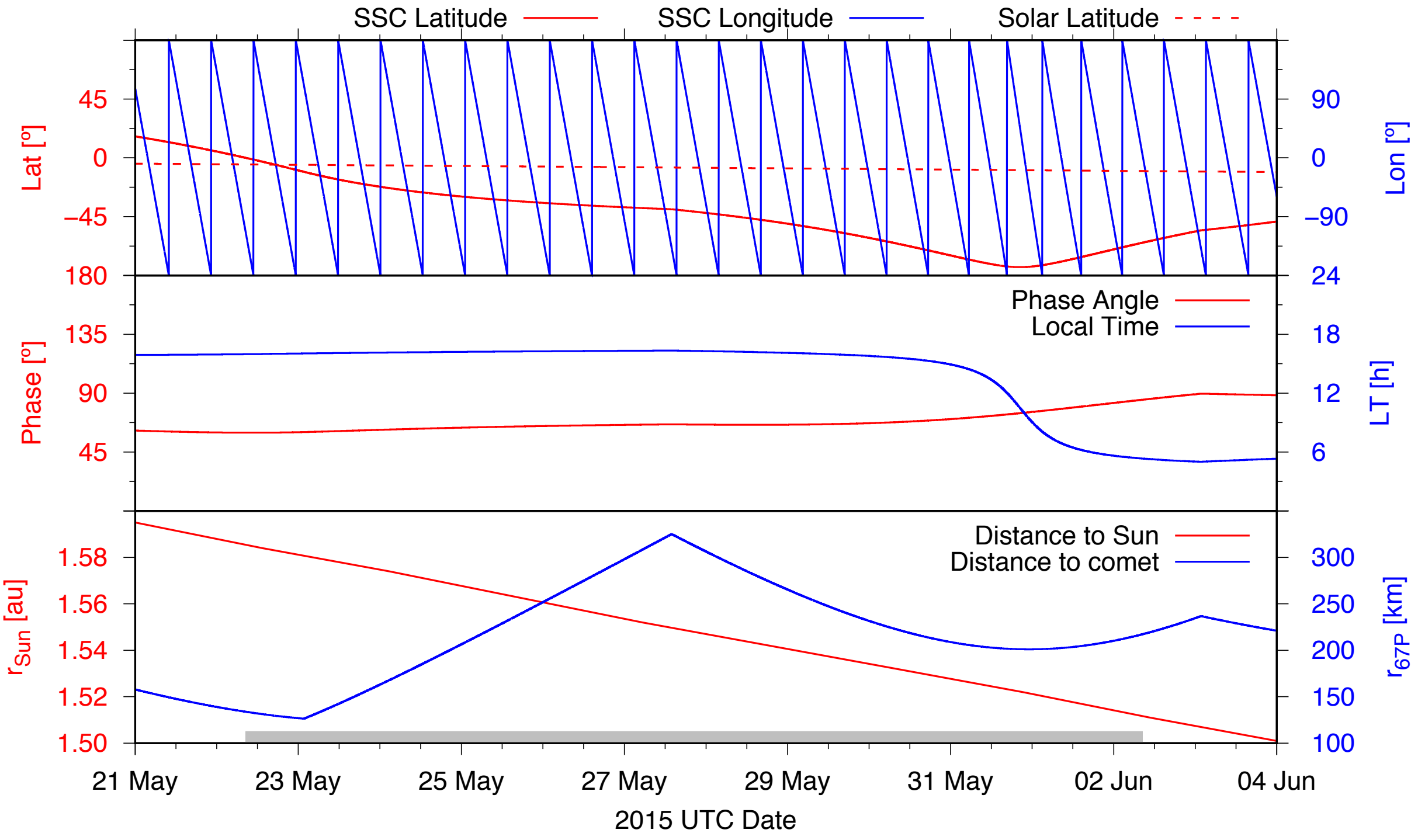

Figure 2: Observation geometry with heliocentric and cometocentric distances in the bottom panel, phase angle and local time in the middle panel, and sub-spacecraft longitude and latitude and sub-solar latitude in the top panel. The investigated period was suitable for deriving bulk abundances in the ices of comet 67P/Churyumov-Gerasimenko (Calmonte et al. 2016) and is indicated by the gray bar at the bottom corresponding to the time when Rosetta was above the dayside of the comet (phase angle < 85°) and above the more active southern summer hemisphere (sub spacecraft latitude < 0°).

Some of the volatile species reported here have also been observed by other scientific instruments on board Rosetta, including $CO_2$, $CH_4$, and OCS by VIRTIS (Bockelée-Morvan et al. 2016) and $H_2O$ by MIRO (Marshall *et al.* 2017). In particular, MIRO measurements (Biver *et al.* 2019) indicate a lower amount of water near perihelion compared to ROSINA-derived water production rates (Hansen et al. 2016). Consequentially, their relative abundances with respect to water may be different. This is in part due to the time period selected for analysis, though several other possible explanations have also been put forward. For instance, distributed sources of volatile species near the nucleus, where a portion of the volatile species are trapped in grains, would be invisible to remote sensing instruments.

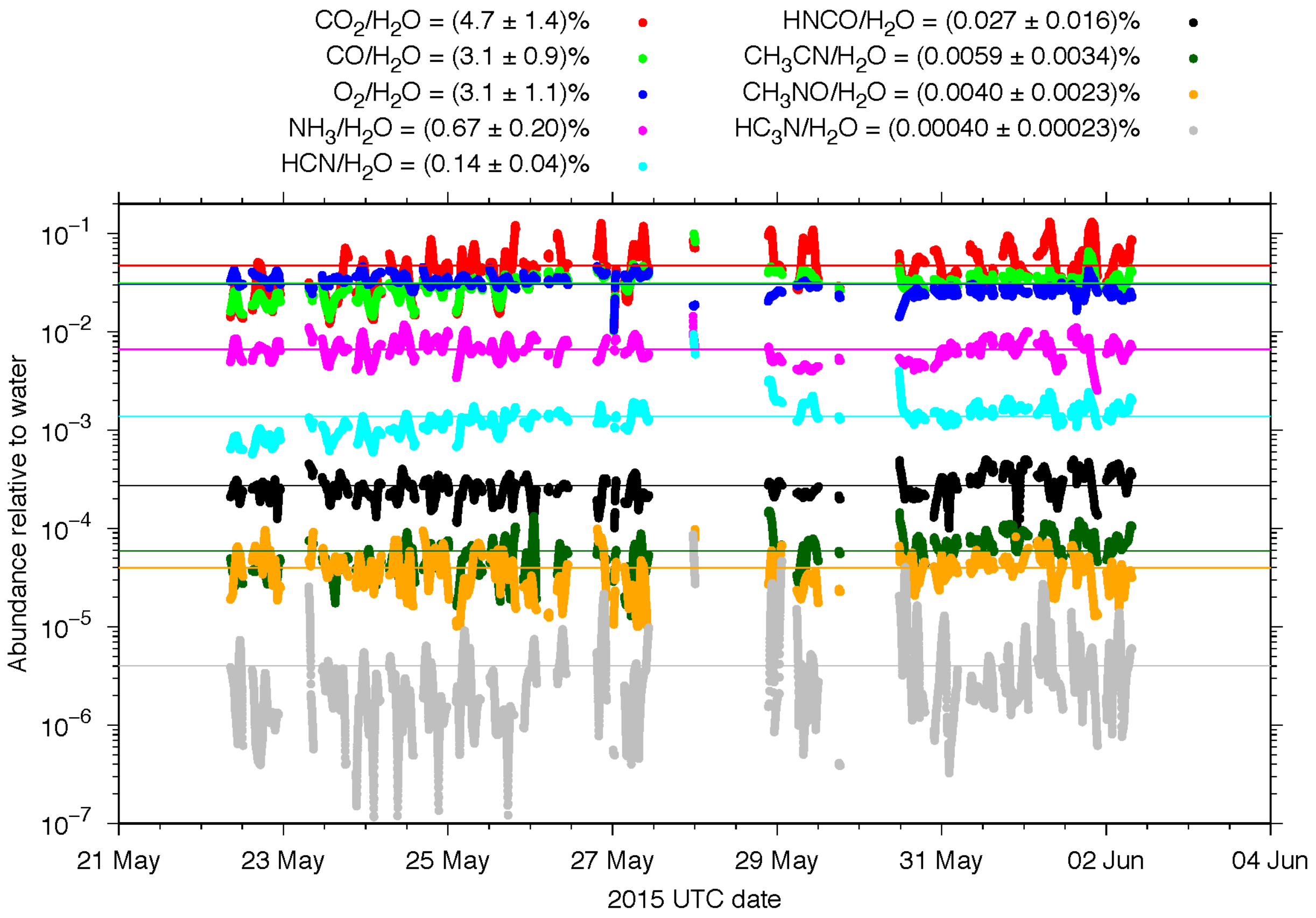

Figure 3: Measured abundances of a suite of volatile species with respect to water from the pre-perihelion period at the end of May 2015, suitable for deriving bulk abundances in the ices of comet 67P/Churyumov-Gerasimenko (Calmonte *et al.* 2016). The horizontal lines denote the averages for the observed period. Indicated errors represent 1-σ.

While some of the volatile species have indeed been shown to be associated with grains (De Keyser *et al.* 2017, Altwegg et al. 2016), no clear indication for distributed sources was found for water, and Biver et al. (2019) estimated an upper limit of 50% for contributions from grains. For our analysis, we assumed that grains moving at ~1 m/s would have lost their volatiles by the time they reached Rosetta (Lien 1990), which was more than 100 km from the nucleus. This is in line with numerical models of the major volatile species $H_2O$, $CO_2$, CO,

Table 2: Deduced bulk composition of the ices in comet 67P/Churyumov-Gerasimenko normalized to water based on the May 2015 period. The hydrogen halide abundances have been derived from F/O, Cl/O, and Br/O ratios from Dhooghe *et al.* (2017) and elemental O-abundances in $H_2O$, $CO_2$, CO, and $O_2$. Calibration values have been used for the indicated component. Additional possible isomers are listed in brackets, as a unique identification is not possible with DFMS. Examples include HCN versus HNC and $CH_3CH_2SH$ versus $CH_3SCH_3$. Phosphorus has been identified in the form of PO. However, contributions in other forms are possible (see text). The indicated errors represent 1-σ, except for the hydrogen halides where observed ranges were reported. Also given are the abundances measured above the nothern and southern hemispheres at 3.1 au inbound from Le Roy *et al.* (2015). References: a) Le Roy et al. (2015), b) Bieler *et al.* (2015), c) Rubin *et al.* (2015a), d) Balsiger *et al.* (2015), e) this work, f) Schuhmann *et al.* (2019), g) Schuhmann *et al.* (in preparation), h) Rubin *et al.* (2018), i) Calmonte et al. (2016), j) Dhooghe et al. (2017), k) Rivilla *et al.* (in preparation).

| **Molecule** | **Deduced bulk abundance** | **Northern vs southern hemispheres at 3.1 au** [a)] | **Name (including some isomers)** | **Reference & notes** |
|---|---|---|---|---|
| $H_2O$ | 100 | 100 / 100 | Water | definition |
| $CO_2$ | 4.7 ± 1.4 | 2.5 / 80 | Carbon dioxide | [e)], cf. Figure 3 |
| CO | 3.1 ± 0.9 | 2.7 / 20 | Carbon monoxide | [e)], cf. Figure 3 |
| $O_2$ | 3.1 ± 1.1 | 3.80 ± 0.85 [b)] | Molecular oxygen | [e)], cf. Figure 3 |
| $CH_4$ | 0.34 ± 0.07 | 0.13 / 0.56 | Methane | [f)] |
| $C_2H_6$ | 0.29 ± 0.06 | 0.32 / 3.3 | Ethane | [f)] |
| $C_3H_8$ | 0.018 ± 0.004 | | Propane | [f)] |
| $C_6H_6$ | 0.00069 ± 0.00014 | | Benzene | [f)] |
| $C_7H_8$ | 0.0062 ± 0.0012 | | Toluene | [f)] |
| $CH_3OH$ | 0.21 ± 0.06 | 0.31 / 0.55 | Methanol | [g)] |
| $C_2H_5OH$ | 0.039 ± 0.023 | | Ethanol (Dimethyl ether) | [g)] |
| $CH_3CH_2CHO$ | 0.0047 ± 0.0024 | | Propanal (Acetone) (Propylene oxide) | [g)] |
| $H_2CO$ | 0.32 ± 0.10 | 0.33 / 0.53 | Formaldehyde | [g)] |
| $CH_3CHO$ | 0.047 ±0.017 | 0.01 / 0.024 | Acetaldehyde (Ethylene oxide) (Vinyl alcohol) | [g)] |
| HCOOH | 0.013 ± 0.008 | 0.008 / 0.03 | Formic acid | [g)] |
| $CH_3COOH$ | 0.0034 ± 0.0020 | 0.004 / 0.023 | Acetic acid (Methyl formate) (Glycolaldehyde) | [g)] |
| $(CH_2OH)_2$ | 0.011 ± 0.007 | 0.0008 / < 0.0025 | Ethylene glycol (Methoxymethanol) | [g)] |
| $CH_3(CH_2)_2CHO$ | 0.010 ± 0.003 | | Butanal (Ethoxyethene) | [g)] |
| $CH_3COOCH_3$ | 0.0021 ± 0.0007 | | Methyl acetate (Propanoic acid) (Acetol) | [g)] |
| $NH_3$ | 0.67 ± 0.20 | 0.06 / 0.15 | Ammonia | [e)], cf. Figure 3 |
| $N_2$ | 0.089 ± 0.024 | 0.015 to 0.114 [a), c)] | Molecular nitrogen | [h)] |
| HCN | 0.14 ± 0.04 | 0.09 / 0.62 | Hydrogen cyanide (Hydrogen isocyanide) | [e)], cf. Figure 3 |
| HNCO | 0.027 ± 0.016 | 0.016 / 0.031 | Isocyanic acid (Cyanic acid) | [e)], cf. Figure 3 |
| $CH_3NO$ | 0.0040 ± 0.0023 | <0.0001 / 0.001 | Formamide (Formaldoxime) (Oxaziridine) | [e)], cf. Figure 3 |
| $CH_3CN$ | 0.0059 ± 0.0034 | 0.006 / 0.016 | Acetonitrile (Isocyanomethane) | [e)], cf. Figure 3 |
| $HC_3N$ | 0.00040 ± 0.00023 | < 0.00002 / < 0.0005 | Cyanoacetylene (Isocyanoacetylene) | [e)], cf. Figure 3 |

Table 2 cont.

| Molecule | Deduced bulk abundance | Northern vs southern hemispheres at 3.1 au [a)] | Name (of some isomers) | Reference & notes |
|---|---|---|---|---|
| $H_2O$ | 100 | 100 / 100 | Water | definition |
| $H_2S$ | 1.10 ± 0.46 | 0.67 / 1.75 | Hydrogen sulfide | i) |
| OCS | $0.041^{+0.082}_{-0.020}$ | 0.017 / 0.098 | Carbonyl sulfide | i) |
| SO | $0.071^{+0.142}_{-0.037}$ | 0.004 / 0.0014 | Sulfur monoxide | i) |
| $SO_2$ | $0.127^{+0.254}_{-0.064}$ | 0.011 / 0.041 | Sulfur dioxide | i) |
| $CS_2$ | $0.0057^{+0.0114}_{-0.0028}$ | 0.003 / 0.024 | Carbonyl sulfide | i) |
| $S_2$ | $0.0020^{+0.0040}_{-0.0010}$ | 0.0004 / 0.0013 | Sulfur dimer | i) |
| $H_2CS$ | $0.0027^{+0.0058}_{-0.0024}$ | | Thioformaldehyde | i) |
| S | 0.46 ± 0.36 | | Atomic sulfur | i) |
| $CH_3SH$ | $0.038^{+0.079}_{-0.028}$ | | Methanethiol | i) |
| $CH_3CH_2SH$<br>$CH_3SCH_3$ | $0.00058^{+0.00123}_{-0.00049}$ | | Ethanethiol<br>Dimethyl sulfide | i) |
| Ar | 0.00058 ± 0.00022 | 0.00012 to 0.0027 [d)] | Argon | h) |
| Kr | 0.000049 ± 0.000022 | | Krypton | h) |
| Xe | 0.000024 ± 0.000011 | | Xenon | h) |
| Ne | < 0.000005 | | Neon | h), upper limit |
| HF | $0.011^{+0.037}_{-0.008}$ | | Hydrogen fluoride | j), range given |
| HCl | $0.014^{+0.045}_{-0.012}$ | | Hydrogen chloride | j), range given |
| HBr | $0.00030^{+0.00053}_{-0.00018}$ | | Hydrogen bromide | j), range given |
| PO | $0.011^{+0.022}_{-0.006}$ | | Phosphorous oxide | k) |

and $O_2$ (Fougere et al. 2016b, Läuter et al. 2019) reproducing ROSINA data within errorbars without the use of any distributed sources.

An alternative approach to estimate the relative bulk abundances of the volatile material in a comet is to integrate the total production of volatile species over the entire apparition and relate it to the total water production. This approach is model-dependent and may not be possible for minor species, which were below DFMS' detection limit for large parts of the mission. Nevertheless, this approach has been carried out by Läuter et al. (2019) for the major molecules $H_2O$, $CO_2$, CO, and $O_2$ by inverting an analytically described outgassing pattern of thousands of surface elements and optimizing their activity to fit the tens of thousands of ROSINA measurements. Over the two years when Rosetta was accompanying the comet, the relative losses were $r_{CO_2} = (5.9 \pm 3.0) \cdot 10^{-2}$, $r_{CO} = (2.3 \pm 1.0) \cdot 10^{-2}$, and $r_{O_2} = (1.6 \pm 0.7) \cdot 10^{-2}$. The corresponding numbers by Combi *et al.* (in preparation) are $r_{CO_2} = 7.4 \cdot 10^{-2}$, $r_{CO} = 2.7 \cdot 10^{-2}$, and $r_{O_2} = 1.6 \cdot 10^{-2}$ and are based on a fully kinetic Direct Simulation Monte Carlo model that represents the molecules and their interaction in the coma by a much smaller set of weighted model particles. These relative numbers may also be seen as an estimate for the bulk abundance of the volatiles in 67P/C-G, assuming that outgassing during the remaining part of 67P/C-G's orbit through its aphelion can be neglected.

Some of the deviations of our results based on the May 2015 period in Table 2 compared to the integrals over the whole mission from Läuter et al. (2019) and Combi et al. (in preparation) can be explained by the different methods used. Late in the mission, at heliocentric distances beyond ~3 au, $CO_2$ became the dominant species as $H_2O$ outgassing decreased sharply (Hansen et al. 2016, Läuter et al. 2019). Consequentially, the integral over the entire mission yields a $CO_2/H_2O$ ratio that is about 25% higher. Nevertheless, the ratios obtained by Läuter et al. (2019) have overlapping error bars with those for the ratios derived for the end of May 2015 period. The three approaches yield a somewhat lower $O_2/H_2O$ compared to $r_{O_2} = (3.80 \pm 0.85) \cdot 10^{-2}$ by Bieler et al. (2015). This difference is a consequence of a decreasing amount of $O_2$ with respect to $H_2O$ towards perihelion (Bieler et al. 2015, Fougere et al. 2016b, Läuter et al. 2019, Combi et al. in preparation).

Table 2 lists a collection of derived bulk abundances in the ices of comet 67P/Churyumov-Gerasimenko, including earlier works on sulphur (Calmonte et al. 2016), halogens (Dhooghe et al. 2017), noble gases, and $N_2$ (Rubin et al. 2018). Furthermore, Rivilla et al. (in preparation) identified phosphorous monoxide, PO, as the main carrier of phosphorus in the comet. Table 2 also lists the deduced abundance of PO, assuming its correlation to SO, derived at a heliocentric distance of 3 au, also applies to May 2015. $PO_x$ with $x \geq 2$ has not been identified in DFMS mass spectra. Only upper limits could be established for $PH_3$ and PN due to interference from $^{18}O^{16}O$ and $CH^{32}S$, respectively. On the other hand, an additional contribution from phosphorus in its elemental form could not be excluded.

Several of the components listed in Table 2 may have been present in the form of a different isomer, possibly even from another chemical functional group. The exact identification of the parent, however, cannot be derived with DFMS, e.g., in the case of hydrogen cyanide versus hydrogen isocyanide (HCN versus HNC, Le Roy et al. (2015)) and ethanethiol versus dimethyl sulphide ($CH_3CH_2SH$ versus $CH_3SCH_3$, Calmonte et al. (2016)). The identification of parent molecules based on the different fragmentation patterns is also very difficult due to the high number of molecules contributing to an even larger suite of daughter species and the presence of a large fraction of unsaturated CH-, CHO-, CHS-, and CHON-bearing species (Altwegg *et al.* 2017b, Schuhmann et al. 2019). To complicate matters further, a mix of the different isomers may also be present at the same time. As a consequence of this, Table 2 lists which parent was assumed when deriving abundances with the corresponding calibration values from Table 1. However, it is possible that a component is fully or partially present in the form of different isomers and hence some of the possibilities are listed in brackets. Comparing ionization cross-sections and, in particular, fragmentation patterns among the different isomers from NIST and laboratory calibrations, differences of up to 50% in the abundances may result from the lack of precise identification of the corresponding molecule(s).

Table 3: Elemental abundances in the ices of comet 67P/Churyumov-Gerasimenko based on Table 2 (left) and in the refractories (right) measured by COSIMA from Bardyn et al. (2017). All ratios are normalized to oxygen. Note: H abundance in the refractories (right column) is estimated (Bardyn et al. 2017).

| Element | Ratio volatiles | Ratio refractories |
|---|---|---|
| H/O | $1.75^{+0.11}_{-0.10}$ | $0.99^{+1.14}_{-0.57}$ |
| O/O | 1.0 (definition) | 1.0 (definition) |
| C/O | $8.29^{+2.64}_{-2.53} \cdot 10^{-2}$ | $1.00^{+1.75}_{-0.94}$ |
| S/O | $1.55^{+1.19}_{-0.83} \cdot 10^{-2}$ | |
| N/O | $8.51^{+2.66}_{-2.65} \cdot 10^{-3}$ | $3.47^{+7.05}_{-3.47} \cdot 10^{-2}$ |
| Cl/O | $1.19^{+3.77}_{-0.99} \cdot 10^{-4}$ | |
| P/O | $0.92^{+1.84}_{-0.46} \cdot 10^{-4}$ | |
| F/O | $0.88^{+3.08}_{-0.69} \cdot 10^{-4}$ | |
| Ar/O | $4.84^{+1.86}_{-1.86} \cdot 10^{-6}$ | |
| Br/O | $2.48^{+4.46}_{-1.49} \cdot 10^{-6}$ | |
| Kr/O | $4.09^{+1.85}_{-1.85} \cdot 10^{-7}$ | |
| Xe/O | $2.00^{+0.93}_{-0.92} \cdot 10^{-7}$ | |
| Ne/O | $< 4.17^{+0.24}_{-0.23} \cdot 10^{-8}$ | |
| Si/O | | $1.84^{+2.89}_{-1.39} \cdot 10^{-1}$ |
| Fe/O | | $5.26^{+6.09}_{-3.05} \cdot 10^{-2}$ |
| Mg/O | | $2.11^{+3.22}_{-1.61} \cdot 10^{-2}$ |
| Na/O | | $1.47^{+3.41}_{-1.27} \cdot 10^{-2}$ |
| Al/O | | $3.21^{+6.23}_{-2.69} \cdot 10^{-3}$ |
| Ca/O | | $1.00^{+1.40}_{-0.72} \cdot 10^{-3}$ |
| Mn/O | | $7.89^{+9.28}_{-4.84} \cdot 10^{-4}$ |
| Cr/O | | $4.89^{+6.54}_{-3.38} \cdot 10^{-4}$ |
| K/O | | $3.89^{+5.79}_{-2.91} \cdot 10^{-4}$ |

Based on the abundances of volatile species presented in Table 2, the elemental abundances in the volatile species of comet 67P/C-G were derived. A summary is presented in Table 3, where all elements have been normalized to oxygen and sorted by their relative abundance in the volatile phase. The most abundant elements are hydrogen and oxygen as a consequence of water being the dominant volatile species in the comet. The S/O ratio takes all O-bearing molecules from Table 2 into account and thus differs slightly from $S/O = 1.47 \cdot 10^{-2}$ by Calmonte et al. (2016) based on $H_2O$, $CO_2$, CO, and $O_2$ for the period between the equinox and perihelion. The same also applies to the F/O, Cl/O, and Br/O ratios which were derived in Dhooghe et al. (2017) based on the main O-bearing molecules $H_2O$, $CO_2$, CO, and $O_2$.

As discussed above, the mass range of DFMS does not include mass/charge m/Q = 2 u/e, thus the contribution of molecular hydrogen as a cometary parent species has not been accounted for in these calculations. In the case of comet C/2001 A2 (LINEAR), however, Feldman *et al.* (2002) showed that the observed $H_2$ in the coma can be satisfactorily accounted for by photodissociated $H_2O$.

Bardyn et al. (2017) presented elemental abundances in the refractory phase of comet 67P/C-G. From COSIMA measurements, the presence of high abundances of organic molecules was inferred (Fray et al. 2016), where 45% by weight of the analysed matter share similarities with the Insoluble Organic Matter (IOM) found in carbonaceous chondrites. The resulting ratios, normalized to oxygen, have been reproduced in Table 3 for comparison. A subset of these refractory species was also measured with ROSINA: Wurz *et al.* (2015) reported elemental abundances of the sputtered refractory species Na, K, and Ca relative to Si. Early in the mission, the solar wind still reached parts of the surface of nucleus, whereas above other areas, increased activity led to collisions in the coma gas and the attenuation of the solar wind (Wurz et al. 2015, Fuselier *et al.* 2015). Nevertheless, the Na/Si, K/Si, and Ca/Si ratios measured by both COSIMA and ROSINA are in agreement.

Combining the elemental ratios from both the refractory and the volatile phases together with the dust-to-ice ratio (D/I) inside the nucleus provides the elemental ratios in the comet. This shall be discussed in the next section.

## 4. Discussion and conclusions

The coma of comet 67P/C-G contains a whole suite of low-temperature volatile species. The measured ratios among these molecules have been used to derive the structure and the formation temperature of the ices in comet 67P/C-G (Rubin et al. 2015a). From the $N_2/CO = (5.70 \pm 0.66)$‰ measured early in the mission at ~3.1 au, Rubin et al. (2015a) derived a formation temperature between 24 and 27 K based on laboratory measurements of the relative trapping efficiencies of $N_2$ versus CO in amorphous water ice at different temperatures (Bar-Nun *et al.* 2007). The case of a crystalline origin of the cometary ices, i.e., the reconfiguration of the ices in the protosolar nebula, was studied by Lectez *et al.* (2015) and yielded a somewhat higher formation temperature of around 45 K. Taking the $Ar/CO = (0.187 \pm 0.091)$‰ ratio into account, Mousis *et al.* (2018) then suggested a further refined crystalline ice model requiring multiple layers of reconfigured ices.

Table 2 yields $N_2/CO = (2.9 \pm 1.2) \cdot 10^{-2}$ in the bulk of the nucleus for the May 2015 period, which is about 5 times higher than the value derived in Rubin et al. (2015a) for the October 2014 period listed above. This result hints at an even lower formation temperature, closer to 20 K for the amorphous ice case. Such a temperature would also be consistent with the temperature range derived independently by Taquet *et al.* (2016) from the measured $O_2/H_2O = (3.80 \pm 0.85) \cdot 10^{-2}$ in comet 67P/C-G by Bieler et al. (2015), which is compatible within error bars with the value derived here $O_2/H_2O = (3.1 \pm 1.1) \cdot 10^{-2}$ from Table 2.

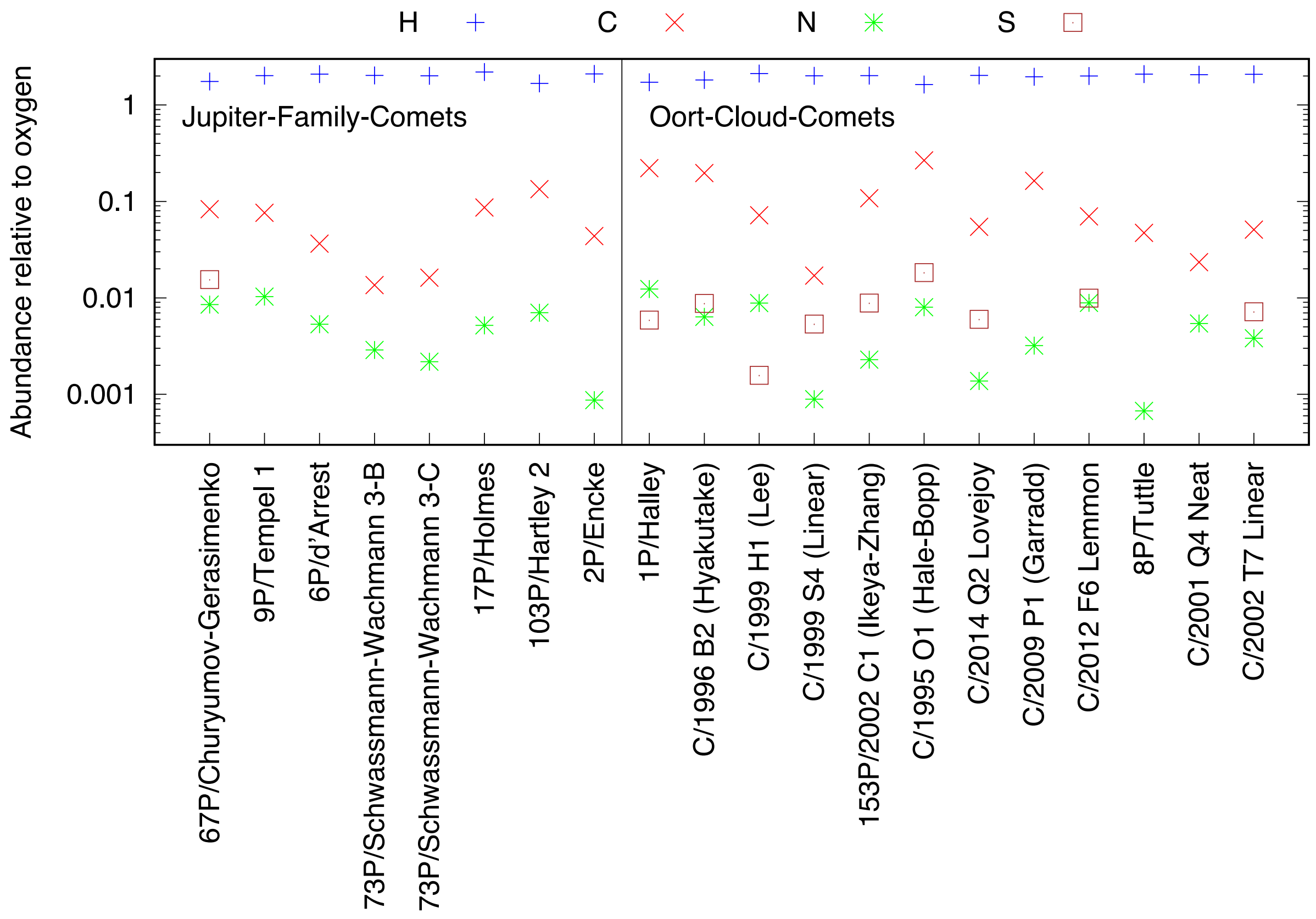

Figure 4: Elemental abundances in the volatile species normalized to oxygen for various comets. Data are taken from Feldman *et al.* (1987), Eberhardt (1999), and Rubin *et al.* (2015b) for 1P/Halley; Biver *et al.* (1999) for Hyakutake; Bockelée-Morvan *et al.* (2004) and references therein for Lee, Ikeya-Zhang, Hale-Bopp, S4 Linear, and Hyakutake; DiSanti *et al.* (2006) for T7 Linear; Remijan *et al.* (2006) and Biver *et al.* (2007) for Neat and T7 Linear; Böhnhardt *et al.* (2008), Bonev *et al.* (2008), and Kobayashi *et al.* (2010) for Tuttle; Paganini *et al.* (2012) for Garradd; Fougere *et al.* (2013) for $CO_2$ estimate in Hartley 2; Biver *et al.* (2014) for Hale-Bopp, Lovejoy, and Lemmon; Biver *et al.* (2016) for Lovejoy and Lemmon; and Dello Russo *et al.* (2016) with references therein for Hyakutake, Encke, S4 Linear, Ikeya-Zhang, Tempel 1, d'Arrest, Schwassmann-Wachmann 3, Holmes, Hartley 2, and Lee.

Elemental abundances in the volatile species can be derived for a suite of comets. Figure 4 shows a comparison of the major elements H, C, N, and S normalized to O in cometary volatile species for both, Oort Cloud Comets (OCCs) and JFCs. While there is some scatter in the observations, i.e., the number of species observed is different from comet to comet, some conclusions can still be drawn. Hydrogen and oxygen are the two major elements in the volatile species, given that water is the main component in cometary ices. The third most abundant element is carbon, originating mostly from CO and $CO_2$, although its relative abundance varies by more than a factor of 10, which is in line with the very variable CO abundances within comets (Dello Russo et al. 2016). Trends are difficult to assess as perihelion distances, which are a measure of peak heating, and heliocentric distances during the times of observation vary among comets (A'Hearn *et al.* 2012). The same applies to the number of perihelion passages a comet has undergone in the past. Furthermore, the abundance of $CO_2$ is unknown for many comets, but may contribute substantially to the C-content in the ices. The abundance of S-bearing and N-bearing species is also quite variable amongst comets, with the caveat that there is very little information available on S-bearing species for JFCs apart from 67P/C-G. Nonetheless, this supports A'Hearn et al. (2012) who found no systematic differences between the OCCs and JFCs.

Table 4: Elemental H, C, N, and O abundances in the ice (Table 3) and dust (Bardyn *et al.* (2017) and Fray *et al.* (2016) with estimated H/Si) in comet 67P/Churyumov-Gerasimenko per silicon atom for dust-to-ice ratios of 1 and 3 by weight.

| Element | Dust | Volatiles (D/I = 1) | Volatiles (D/I = 3) | Combined (D/I = 1) | Combined (D/I = 3) |
|---|---|---|---|---|---|
| H/Si | $5.4^{+5.7}_{-2.6}$ | $18.5^{+0.3}_{-0.3}$ | $6.2^{+0.1}_{-0.1}$ | $24^{+6}_{-3}$ | $12^{+6}_{-3}$ |
| C/Si | $5.4^{+9.2}_{-4.8}$ | $0.87^{+0.27}_{-0.26}$ | $0.29^{+0.09}_{-0.09}$ | $6.3^{+9.2}_{-4.8}$ | $5.7^{+9.2}_{-4.8}$ |
| N/Si | $0.19^{+0.37}_{-0.19}$ | $0.090^{+0.028}_{-0.028}$ | $0.030^{+0.009}_{-0.009}$ | $0.28^{+0.37}_{-0.19}$ | $0.22^{+0.37}_{-0.19}$ |
| O/Si | $5.4^{+8.5}_{-4.1}$ | $10.5^{+0.6}_{-0.6}$ | $3.5^{+0.2}_{-0.2}$ | $16^{+9}_{-4}$ | $8.9^{+8.5}_{-4.1}$ |

The elemental abundances in the bulk of comet 67P/C-G depend on both the dust-to-ice (D/I) ratio (by weight) and the elemental composition in both the dust and the gas (cf. Table 3). Table 4 lists the ratios of the elements abundantly found in the volatile species relative to Si, which is typically found in the refractory material. The contributions have been obtained separately for the dust, the volatiles, and the nucleus total for two representative D/I ratios. Estimated dust-to-gas (D/G) ratios in the coma based on ROSINA measurements have been presented by Hansen et al. (2016) and Läuter et al. (2019) and yield a D/I ratio of around 1, even when including some 20% of dry dust backfall (Keller *et al.* 2017), which may be missing in the D/G ratio measured in the coma.

For comparison, we also investigated a higher D/I ratio of 3, within the range of the (D/I =) $D/G = 4 \pm 2$ reported by Rotundi *et al.* (2015). Even higher D/I ratios (cf. Fulle *et al.* (2018)), which would be associated with a higher percentage of backfall of dry dust, can easily be derived from the numbers given in Table 4. The most prominent differences when changing the D/I ratio are expected in the H/Si and O/Si ratios, given that water is the main component in the cometary ices.

Table 5: Distribution of H, C, N, O, and Si by number in grains and ice. Based on Table 3 for two dust-to-ice ratios inside the nucleus, D/I = 1 and D/I = 3 by weight.

| Element | D/I = 1 | | D/I = 3 | |
|---|---|---|---|---|
| | Dust [%] | Ice [%] | Dust [%] | Ice [%] |
| H | $22^{+18}_{-8}$ | $78^{+18}_{-8}$ | $46^{+26}_{-12}$ | $54^{+26}_{-12}$ |
| O | $34^{+35}_{-17}$ | $66^{+34}_{-17}$ | $61^{+37}_{-18}$ | $39^{+37}_{-18}$ |
| C | $86^{+14}_{-11}$ | $14^{+20}_{-11}$ | $95^{+5}_{-4}$ | $5^{+8}_{-4}$ |
| N | $68^{+32}_{-23}$ | $32^{+44}_{-23}$ | $86^{+14}_{-12}$ | $14^{+24}_{-12}$ |
| Si | 100 | 0 | 100 | 0 |

As a consequence of these numbers, the distribution of the different elements amongst the dust and the ice could also be computed. Table 5 shows that the majority of the hydrogen atoms resides in the ices. However, this changes for high D/I ratios. A similar picture arises for oxygen, whereas carbon and nitrogen seem to be predominantly in the refractories. These results are consistent with earlier findings at OCC 1P/Halley by Geiss (1987), who assumed that the dust-to-gas ratio in the coma reflected the dust-to-ice ratio in the nucleus, i.e., D/G = D/I = 0.5 based on McDonnell *et al.* (1986). The inferred high abundance of C-bearing species in the refractory material of 67P/C-G derived from COSIMA measurements (Fray et al. 2016, Bardyn et al. 2017) are in agreement with measurements obtained at comet 1P/Halley (Jessberger *et al.* 1988). However, the macromolecular matter observed by COSIMA and the zoo of (semi-)volatile organic molecules detected by ROSINA (Altwegg et al. 2017b, Schuhmann et al. 2019) still leaves room for a suite of complex organics to fill the gap between the two. Whether such a continuum exists is still an open question and could further affect the distribution of the elements H, O, C, and N between dust and ice.

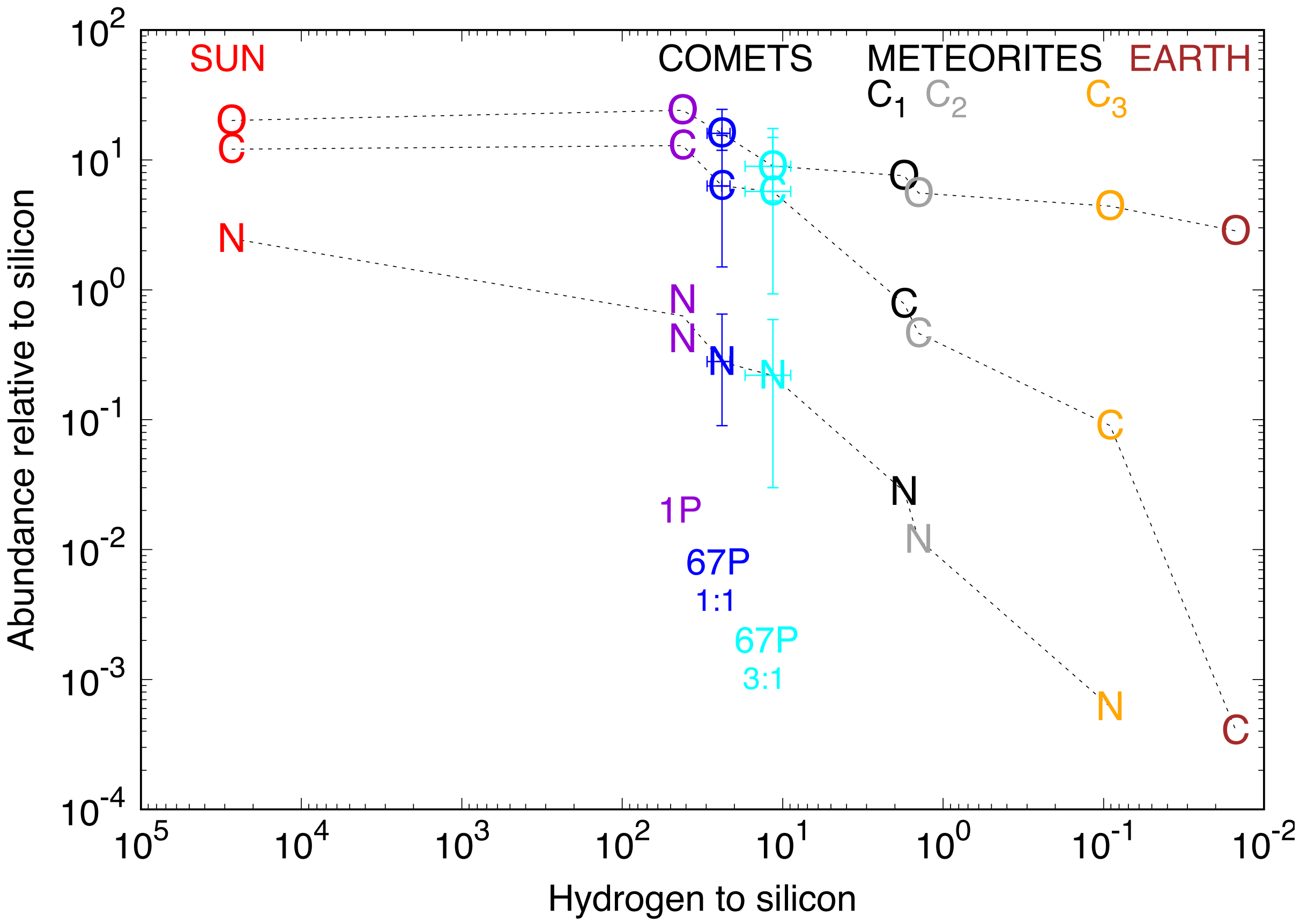

Figure 5: Elemental abundances of H, C, N, and O relative to Si in the solar nebula (red; Anders and Ebihara (1982)), comet 1P/Halley (purple, for D/I = 0.5 and two estimates for N from Balsiger *et al.* (1986) and Allen *et al.* (1987); Geiss (1987)), comet 67P/C-G (for D/I=1 in blue and D/I=3 in turquoise), carbonaceous chondrites ($C_1$ in black, $C_2$ in grey, and $C_3$ in orange; Kerridge (1985)), and an estimate of the Earth's crust and mantle (brown; Larimer (1971), Wänke (1981), and Wänke *et al.* (1984)). See Table 4 for data on 67P/C-G and Geiss (1987) for the others.

The combined elemental abundance ratios from Table 4 (cf. Bardyn et al. (2017), Fray *et al.* (2017)) can also be compared to other objects in the solar system. Figure 5 reproduces the plot by Geiss (1987) with the addition of the comet 67P/C-G. Similar to 1P/Halley, near-solar O/Si and C/Si ratios were found in 67P/C-G and N/Si was depleted compared to solar. D/I ratios inside the nucleus versus in the coma are still a matter of debate, fuelled by the redisposition of material at 67P/C-G (Rubin *et al.* 2014, Thomas *et al.* 2015). Both comets hence display consistent ratios considering the error bars and the uncertainty in the D/I ratio, the latter impacting mostly H/Si and O/Si as discussed earlier. Meteorites, and even more so the Earth, show higher degrees of devolatilization. In particular, both C/Si and N/Si are much lower compared to both the Sun and comets.

The depletion of H/Si in the ices of the comet compared to the Sun was expected due to the high volatility of H (Geiss 1987). The uncertainty of N-abundance in 1P/Halley was larger than for the other elements, hence two estimates were provided. The Rosetta mission obtained a better estimate of the total volatile N-abundance through the detailed quantification of the amount of $N_2$ (Rubin et al. 2015a). The major contributor to the elemental abundance of nitrogen in the volatile species is $NH_3$, followed by $N_2$ and HCN, the latter including HNC. Several additional N-bearing species have been identified, though the N-depletion in comet 67P/C-G, as shown in Figure 5, is still comparable to earlier measurements at comet 1P/Halley.

Other differences between OCC 1P/Halley and JFC 67P/C-G can be explained by the distinct D/I ratios and the lower CO abundance derived for the ices of 67P/C-G in comparison with the >10% of CO with respect to water at 1P/Halley (Eberhardt 1999, Rubin *et al.* 2011). Furthermore, their different dynamic histories may have led to additional fractionation, as 67P/C-G was subjected to heating during the transition through the Centaur stage (Guilbert-Lepoutre *et al.* 2016).

Potential differences between the comet families, JFCs versus OCCs, have also been discussed based on the assumption that the different families formed in two distinct zones (Duncan & Levison 1997). However, later results based on D/H measurements (cf. Altwegg *et al.* (2017a) and references therein) in the water of a suite of OCCs and JFCs support the theory that the comets in both families may have formed over a wide range of overlapping heliocentric distances before being scattered to where they are found today (Brasser & Morbidelli 2013). Based on the results shown in Figure 5, Geiss (1987) concluded that comets are regular members of our solar system and that they must have preserved the original material better than other objects. This certainly also holds for JFC 67P/C-G.

## Acknowledgements

ROSINA would not have produced such outstanding results without the work of the many engineers, technicians, and scientists involved in the mission, in the Rosetta spacecraft team, and in the ROSINA instrument team over the last 20 years, whose contributions are gratefully acknowledged. Rosetta is an ESA mission with contributions from its member states and NASA. We acknowledge herewith the work of the whole ESA Rosetta team.

Work at the University of Bern was funded by the State of Bern, the Swiss National Science Foundation (SNSF, 200021_165869 and 200020_182418), the Swiss State Secretariat for Education, Research and Innovation (SERI) under contract number 16.0008-2, and by the European Space Agency's PRODEX Programme. MND acknowledges the financial support of the SNSF Ambizione grant 180079, the Center of Space and Habitability (CSH) Fellowship, and the International Astronomical Union (IAU) Gruber Foundation Fellowship. SFW acknowledges the financial support of the SNSF Eccellenza Professorial

Fellowship PCEFP2_181150. JDK acknowledges support by the Belgian Science Policy Office via PRODEX/ROSINA PEA 90020. JHW acknowledges JPL subcontract NAS703001TONMO710889. SAF acknowledges JPL contract 1496541. Work at MPS was funded by the Max-Planck Society and BMWI under contract 50QP1302. Work at UoM was supported by contracts JPL 1266313 and JPL 1266314 from the US Rosetta Project.

## Bibliography

A'Hearn M., F. , et al., 2012, Astrophys. J. Lett., 758, 29
Allen M., et al., 1987, Astron Astrophys, 187, 502
Altwegg K., et al., 2016, Sci Adv, 2, e1600285
Altwegg K., et al., 2017a, Philosophical Transactions of the Royal Society A: Mathematical, Physical and Engineering Sciences, 375
Altwegg K., et al., 2017b, Monthly Notices of the Royal Astronomical Society, 469, S130
Anders E., Ebihara M., 1982, Geochimica et Cosmochimica Acta, 46, 2363
Balsiger H., et al., 2015, Sci Adv, 1, e1500377
Balsiger H., et al., 2007, Sp. Sci. Rev., 128, 745
Balsiger H., et al., 1986, Nature, 321, 330
Bar-Nun A., Notesco G., Owen T., 2007, Icarus, 190, 655
Bardyn A., et al., 2017, Monthly Notices of the Royal Astronomical Society, 469, S712
Bibring J.P., et al., 2015, Science, 349, 493
Bieler A., et al., 2015, Nature, 526, 678
Biver N., et al., 1999, The Astronomical Journal, 118, 1850
Biver N., et al., 2007, Planet. Space Sci., 55, 1058
Biver N., et al., 2014, Astron. Astrophys., 566
Biver N., et al., 2019, Astron. Astrophys.,
Biver N., et al., 2016, Astron. Astrophys., 589
Bockelée-Morvan D., et al., 2015, Sp. Sci. Rev., 197, 47
Bockelée-Morvan D., et al., 2016, Monthly Notices of the Royal Astronomical Society, 462, S170
Bockelée-Morvan D., Crovisier J., Mumma M.J., Weaver H.A., Keller H.U., Weaver H.A., 2004, in: Festou M.C., Keller H.U., Weaver H.A. eds., Comets II. Univ. of Arizona Press, Tucson, AZ, p. 391
Bockelée-Morvan D., et al., 2000, Astron. Astrophys., 353, 1101
Bodewits D., et al., 2016, The astronomical journal, 152, 130
Böhnhardt H., Mumma M.J., Villanueva G.L., DiSanti M.A., Bonev B.P., Lippi M., Käufl H.U., 2008, Astrophys. J. Lett., 683, L71
Bonev B.P., Mumma M.J., Radeva Y.L., DiSanti M.A., Gibb E.L., Villanueva G.L., 2008, Astrophys. J. Lett., 680, L61
Brasser R., Morbidelli A., 2013, Icarus, 225, 40
Brownlee D., et al., 2006, Science, 314, 1711
Calmonte U., et al., 2016, Monthly Notices of the Royal Astronomical Society, 462, S253
Calmonte U.M., 2015, PhD, Universität Bern
Combi M., et al., in preparation, Icarus,
De Keyser J., et al., 2017, Monthly Notices of the Royal Astronomical Society, 469, S695
Dello Russo N., Kawakita H., Vervack R.J., Weaver H.A., 2016, Icarus, 278, 301
Dhooghe F., et al., 2017, Monthly Notices of the Royal Astronomical Society, 472, 1336

DiSanti M.A., Bonev B.P., Magee-Sauer K., Russo N.D., Mumma M.J., Reuter D.C., Villanueva G.L., 2006, Astrophys. J. Lett., 650, 470
Duncan M.J., Levison H.F., 1997, Science, 276, 1670
Eberhardt P., 1999, Sp. Sci. Rev., 90, 45
Feldman P.D., et al., 2015, Astron. Astrophys., 583
Feldman P.D., et al., 1987, Astron. Astrophys., 187, 325
Feldman P.D., Weaver H.A., Burgh E.B., 2002, Astrophys. J. Lett., 576, L91
Fougere N., et al., 2016a, Astron. Astrophys., 588, A134
Fougere N., et al., 2016b, Monthly Notices of the Royal Astronomical Society, 462, S156
Fougere N., Combi M.R., Rubin M., Tenishev V., 2013, Icarus, 225, 688
Fray N., et al., 2016, Nature, 538, 72
Fray N., et al., 2017, Monthly Notices of the Royal Astronomical Society, 469, S506
Fulle M., et al., 2018, Monthly Notices of the Royal Astronomical Society, 482, 3326
Furuya K., van Dishoeck E.F., Aikawa Y., 2016, Astron. Astrophys., 586, A127
Fuselier S.A., et al., 2015, Astron. Astrophys., 583
Gasc S., et al., 2017a, Monthly Notices of the Royal Astronomical Society, 469, S108
Gasc S., et al., 2017b, Planet. Space Sci., 135, 64
Geiss J., 1987, Astron. Astrophys., 187, 859
Glassmeier K.H., Boehnhardt H., Koschny D., Kuhrt E., Richter I., 2007, Sp. Sci. Rev., 128, 1
Goesmann F., et al., 2015, Science, 349, aab0689
Graf S., et al., 2004, J Geophys Res-Planet, 109
Guilbert-Lepoutre A., Rosenberg E.D., Prialnik D., Besse S., 2016, Monthly Notices of the Royal Astronomical Society, 462, S146
Hansen K.C., et al., 2016, Monthly Notices of the Royal Astronomical Society, 462, S491
Hässig M., et al., 2015, Science, 347, aaa0276
Herbst E., van Dishoeck E.F., 2009, Annual Review of Astronomy and Astrophysics, 47, 427
Hoang M., et al., 2017, Astron. Astrophys., 600
Jessberger E.K., Christoforidis A., Kissel J., 1988, Nature, 332, 691
Keller H.U., et al., 2017, Monthly Notices of the Royal Astronomical Society, 469, S357
Kerridge J.F., 1985, Geochim Cosmochim Acta, 49, 1707
Kim Y.-K., et al., 2005.
Kissel J., et al., 2007, Sp. Sci. Rev., 128, 823
Kobayashi H., et al., 2010, Astron. Astrophys., 509, A80
Larimer J.W., 1971, Geochimica et Cosmochimica Acta, 35, 769
Läuter M., Kramer T., Rubin M., Altwegg K., 2019, Monthly Notices of the Royal Astronomical Society, 483, 852
Le Roy L., et al., 2015, Astron. Astrophys., 583, A1
Lectez S., Simon J.M., Mousis O., Picaud S., Altwegg K., Rubin M., Salazar J.M., 2015, Astrophys. J. Lett., 805, L1
Lien D.J., 1990, Astrophys. J., 355, 680
Marschall R., et al., 2017, Astron. Astrophys., 605, A112
Marshall D.W., et al., 2017, Astron. Astrophys., 603
Marty B., et al., 2017, Science, 356, 1069
Mattauch J., Herzog R.F.K., 1934, Z. Physik, 9, 786
McDonnell J., Kissel J., Grün E., Grard R., Langevin Y., Olearczyk R., Perry C., Zarnecki J., 1986, ESLAB Symposium on the Exploration of Halley's Comet.
McKeegan K.D., et al., 2006, Science, 314, 1724

Mousis O., et al., 2018, Astrophys. J. Lett., 865, L11
Nevejans D., Neefs E., Kavadias S., Merken P., Van Hoof C., 2002, International Journal of Mass Spectrometry, 215, 77
Paganini L., Mumma M., Villanueva G., DiSanti M., Bonev B., Lippi M., Boehnhardt H., 2012, Astrophys. J. Lett., 748, L13
Remijan A.J., et al., 2006, Astrophys. J. Lett., 643, 567
Rivilla V., Altwegg K., Dishoeck E.F.v., Drozdovskaya M., al. e., in preparation, Monthly Notices of the Royal Astronomical Society,
Rotundi A., et al., 2015, Science, 347, aaa3905
Rubin M., et al., 2015a, Science, 348, 232
Rubin M., et al., 2018, Sci Adv, 4, eaar6297
Rubin M., Altwegg K., van Dishoeck E.F., Schwehm G., 2015b, Astrophys. J. Lett., 815, L11
Rubin M., Fougere N., Altwegg K., Combi M.R., Le Roy L., Tenishev V.M., Thomas N., 2014, Astrophys. J., 788, 168
Rubin M., Tenishev V.M., Combi M.R., Hansen K.C., Gombosi T.I., Altwegg K., Balsiger H., 2011, Icarus, 213, 655
Schroeder I., et al., 2018, Astron. Astrophys.,
Schuhmann M., et al., in preparation, ACS Earth and Space Chemistry,
Schuhmann M., et al., 2019, Astron. Astrophys.,
Schwehm G., Schulz R., 1999, Sp. Sci. Rev., 90, 313
Sierks H., et al., 2015, Science, 347, aaa1044
Taquet V., Furuya K., Walsh C., van Dishoeck E.F., 2016, Monthly Notices of the Royal Astronomical Society, 462, S99
Tenishev V., Combi M., Davidsson B., 2008, Astrophys. J., 685, 659
Thomas N., et al., 2015, Astron. Astrophys., 583
Tzou C.-Y., 2017, PhD PhD thesis, University of Bern
Wänke H., 1981, Phil. Trans. R. Soc. Lond. A, 303, 287
Wänke H., Dreibus G., Jagoutz E., 1984, Archaean geochemistry. Springer, p. 1
Wright I.P., Sheridan S., Barber S.J., Morgan G.H., Andrews D.J., Morse A.D., 2015, Science, 349, aab0673
Wurz P., et al., 2015, Astron. Astrophys., 583
Zakharov V.V., Crifo J.F., Rodionov A.V., Rubin M., Altwegg K., 2018, Astron. Astrophys., 618